\documentclass[
 preprint,
 amsmath,amssymb,
 aps,
prb,superscriptaddress
]{revtex4-2}

\usepackage{graphicx} 
\usepackage{dcolumn} 
\usepackage{bm} 
\usepackage{siunitx}
\usepackage{lipsum}

\newcommand{\beginsupplement}{%
        \setcounter{table}{0}
        \renewcommand{\thetable}{S\arabic{table}}%
        \setcounter{figure}{0}
        \renewcommand{\thefigure}{S\arabic{figure}}%
        \renewcommand{\thesubsection}{\roman{subsection}}
     }

\begin{document}

\title{Probing Laser-Induced Spin-Current Generation in Synthetic Ferrimagnets Using Spin Waves}

\author{Tom Lichtenberg}
\thanks{These authors contributed equally.}
\author{Youri L.W. van Hees}
\thanks{These authors contributed equally.}
\author{Maarten Beens}
\author{Caspar J. Levels} 
\author{Reinoud Lavrijsen}
\affiliation{
 Department of Applied Physics, Eindhoven University of Technology, P.O. Box 513, 5600 MB Eindhoven, The Netherlands}
\author{Rembert A. Duine}
\affiliation{
 Department of Applied Physics, Eindhoven University of Technology, P.O. Box 513, 5600 MB Eindhoven, The Netherlands}
\affiliation{
Institute for Theoretical Physics, Utrecht University, Leuvenlaan 4, 3584 CE Utrecht, The Netherlands
}
\author{Bert Koopmans}
\affiliation{
 Department of Applied Physics, Eindhoven University of Technology, P.O. Box 513, 5600 MB Eindhoven, The Netherlands}
 
\date{\today}

\begin{abstract}
Several rare-earth transition-metal ferrimagnetic systems exhibit all-optical magnetization switching upon excitation with a femtosecond laser pulse.
Although this phenomenon is very promising for future opto-magnetic data storage applications, the role of non-local spin transport in these systems is scarcely understood.
Using Co/Gd and Co/Tb bilayers we isolate the contribution of the rare-earth materials to the generated spin currents by using the precessional dynamics they excite in an adjacent ferromagnetic layer as a probe.
By measuring THz standing spin-waves as well as GHz homogeneous precessional modes, we probe both the high- and low-frequency components of these spin currents.
The low-frequency homogeneous mode indicates a significant contribution of Gd to the spin current, but not from Tb, consistent with the difficulty in achieving all-optical switching in Tb-containing systems.
Measurements on the THz frequency spin waves reveal the inability of the rare-earth generated spin currents to excite dynamics at the sub-ps timescale.
We present modelling efforts using the $s$-$d$ model, which effectively reproduce our results and allow us to explain the behavior in terms of the temporal profile of the spin current. 
\end{abstract}

\maketitle
\section{Introduction}
Over the past few decades, femtosecond (fs) laser excitation of magnetic materials has led to the discovery of a rich collection of physical phenomena.
Among these, single fs laser pulse all-optical switching (AOS) of the magnetization in rare-earth transition-metal ferrimagnetic alloys~\cite{stanciu2007,ostler2012} and multilayers~\cite{lalieu2017aos,aviles2019} appears to be especially promising for future memory applications~\cite{kimel2019review}.
This phenomenon was shown to arise from the transfer of angular momentum between two magnetic sublattices~\cite{radu2011}.
In the same period the fs laser excitation of spin currents, mobile electrons carrying spin angular momentum, has been gaining significant interest from a fundamental perspective~\cite{battiato2010,Choi2014}.

A particularly relevant use for these optically excited spin currents exploits their ability to transfer angular momentum between two ferromagnetic layers. 
This was first demonstrated in an experiment investigating the ultrafast laser induced magnetization dynamics of two ferromagnetic layers separated by a conductive spacer layer~\cite{malinowski2008}. 
A diverse body of research into optically generated spin currents has since arisen~\cite{battiato2010,melnikov2011ultrafast,rudolf2012ultrafast,hurst2018spin,dewhurst2018laser, rouzegar2021laser,jimenez2022transition}.
In recent years it has been shown that novel device functionality can be achieved at the intersection of AOS and optically excited spin currents. 
These works focus on systems where an all-optically switchable ferrimagnetic layer is separated by a spacer layer from a ferromagnetic layer. 
Depending on the precise composition of the layers, either the spin current coming from the ferromagnet can influence the AOS process~\cite{van2020deterministic}, or the ferromagnet can be switched by the spin current coming from the switchable layer~\cite{iihama2018,igarashi2020,remy2020}. 
This last case demonstrates the strength of the spin current generated upon excitation of an all-optically switchable system, and begs the question to what extent this spin current plays a role in the switching mechanism.

Although a great deal of AOS research has been performed on Gd(Fe)Co and Co/Gd systems, recent research indicates that Tb can be used as rare-earth (RE) component instead of Gd~\cite{aviles2019,aviles2020}.
However, in these works AOS has been found only for very specific layer thicknesses, a requirement which is not present in layered ferrimagnets containing Gd~\cite{beens2019comparing}.
Work by Choi et al.\ has shown that spin currents generation in Tb is significantly weaker than in Gd~\cite{choi2018}, hinting at an explanation for the increased difficulty in attaining AOS.
Here the accumulation of spins at the far end of a thick conductive spacer layer was measured, which can lead to a distorted spin current profile due to diffusive electron transport~\cite{Choi2014}.
Moreover, the use of alloys makes it difficult to disentangle the spin current contributions originating from different elements.

In this work, we systematically study spin current generation in synthetic ferrimagnets using the collective spin precession they excite in a neighboring layer as a probe.
The basic experimental concept and the noncollinear magnetic bilayers used in this work are sketched in Fig.\,\ref{fig:1}(a). 
The generation layers are synthetic ferrimagnetic bilayers, consisting of Co which couples antiferromagnetically to either Gd or Tb.
Although these latter materials are paramagnetic at room temperature in bulk, a ferromagnetic phase can be stabilized when interfaced with a ferromagnetic material~\cite{haskel2001enhanced}, which decays exponentially away from the interface.
Upon laser excitation of the out-of-plane magnetized generation layer, a spin current which is spin-polarized in the direction of the local magnetization is excited and injected into a ferromagnetic absorption layer with in-plane magnetization.
There the spins exert a torque on the local magnetic moments and excite both the homogeneous precessional mode~\cite{schellekens2014spinwaves,Choi2014} as well as higher-order inhomogeneous spin waves~\cite{Razdolski2017NanoscaleDynamics,lalieu2017thz}.
By extracting the phase, amplitude and frequency of the FMR mode and THz spin waves, we indirectly study the absorbed spin current~\cite{lichtenberg2022probing}, which is commonly expected to scale with the time derivative of the magnetization ($\text{d}m/\text{d}t$), fulfilling conservation of angular momentum~\cite{Choi2014, lichtenberg2022probing}. 
In this framework, these spin-wave parameters are directly proportional to the corresponding parameters of the Fourier component of the spin-torque pulse at the spin-wave frequency. 
Specifically, we probe the phase of the precessional dynamics to investigate the temporal profile of the spin current, and the amplitude to study the spin-current strength and attenuation. 
Changing the thickness of the RE layer changes both spin-current properties, which leads to significant variations in the measured parameters of the precessions. 
By measuring both the homogeneous FMR mode ($\sim\!10$ GHz) and the first-order inhomogeneous mode ($\sim\!0.5$ THz), the spin current can be studied on two distinct timescales, giving access to both the fast as well as the slow components of the spin current separately.
We corroborate our experiments with an $s$-$d$ model which treats local and non-local spin dynamics in a joint description~\cite{Gridnev2016,Tveten2015,Shin2018,Beens2020}.

Our results on the FMR mode show a strong contribution to the excited spin current from Gd, which is in line with previous work~\cite{iihama2018,choi2018} and confirmed by our modelling efforts.
In contrast, the THz mode cannot be efficiently excited by Gd, hinting at the relatively slow nature of the spin current contribution from Gd.
In the Co/Tb system, both the FMR and THz modes are found to vanish with only a slight addition of Tb, indicating weak spin current generation in Tb but strong spin absorption, consistent with the high spin-orbit coupling (SOC) in this material.
These results shed new light on non-local spin dynamics in highly technologically relevant synthetic ferrimagnets, and help elucidate the role of these processes in all-optical magnetic data recording.

\section{Experimental Details}
\begin{figure}[t!]
\includegraphics[width=8.6 cm]{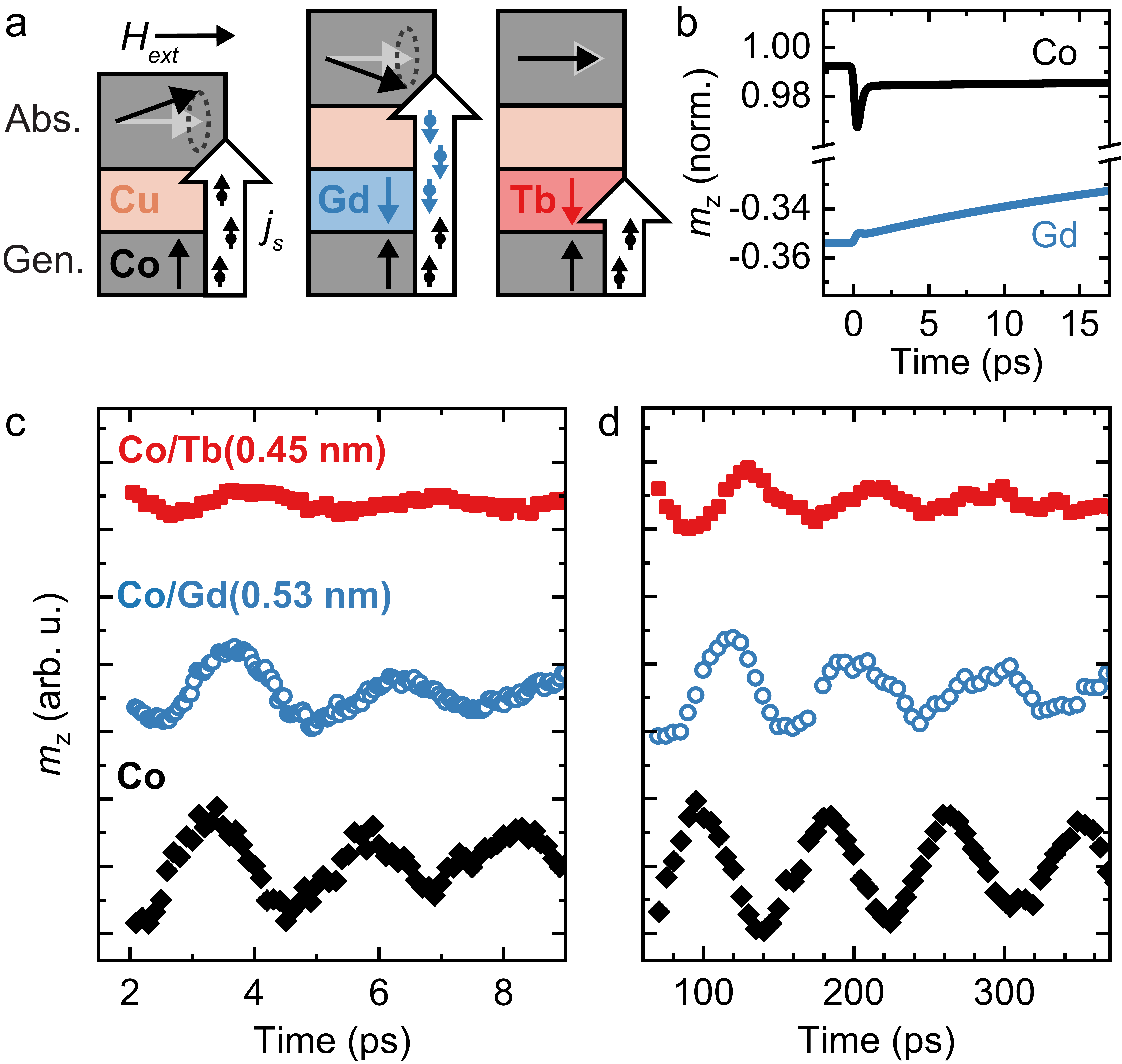}
\caption{\label{fig:1} 
(a) Schematic overview of the experiments, where the effect of the generation layer composition on the response of the absorption layer is sketched for each of the three studied configurations (Co, Co/Gd and Co/Tb). 
(b) Example magnetization response calculated using the \emph{s}-\emph{d} model for Co and for Gd in a Co/Gd bilayer.
(c, d) Response of the absorption layer for each studied configuration. Both (c) the homogeneous ($H_\text{ext}=100$ mT) and (d) the first order inhomogeneous mode ($H_\text{ext}=0$ mT) are plotted, where the latter is measured using \emph{complex MOKE}.
} 
\end{figure}

The noncollinear magnetic bilayer central to this work, as sketched in Fig.\,\ref{fig:1}(a), consists of an out-of-plane magnetized Co/RE generation layer (where RE is either Gd or Tb) and an in-plane magnetized Co absorption layer. 
Spin transport between the two magnetic layers is facilitated by a thin Cu spacer layer, which also serves to magnetically decouple the layers.
The samples are fabricated using DC magnetron sputtering, where we vary the RE layer thickness along the sample length to ensure consistency between measurements.
The full sample stack is given by Si:B(substrate)/{\allowbreak}Ta(4)/{\allowbreak}Pt(4){\allowbreak}/{\allowbreak}Co(1)/{\allowbreak}RE($X$)/{\allowbreak}Cu(2.5){\allowbreak}/{\allowbreak}Co(5){\allowbreak}/{\allowbreak}Pt(2.5) (numbers in brackets indicate the thickness in nm, with $X$ being the variable RE thickness). 

We employ pump-probe spectroscopy to measure the magnetization dynamics upon laser-pulse excitation, using the magneto-optical Kerr effect (MOKE) in the polar configuration to probe the magnetization. 
The sample is placed in a magnetic field with a variable angle of up to 20\textdegree{} with the sample plane.  
A mode-locked Ti:Sapphire laser is used to generate pulses at a repetition rate of 80 MHz and with a wavelength of 780 nm. 
The pulse length at sample position is approximately 150 fs. 
Both pump and probe pulses are focused onto the sample with a spot size of approximately 16 and 8 \SI{}{\micro\metre\squared} respectively, and the pump fluence is about \SI{1}{\milli\joule\per\cm\squared}. 
We use \emph{Complex MOKE} to separate the magnetic contrast of both magnetic layers~\cite{schellekens2014complexmoke}. 
To accurately determine the absolute spin-wave phase in our experiments, the so-called coherence peak is used, which arises at temporal and spatial pump-probe overlap due to interference effects~\cite{eichler1984coherence}. 
For more details on the experimental setup, see Sec.\,I of the Supplementary Information~\cite{Sup}.

\begin{figure*}[t!]
\includegraphics[width=\linewidth]{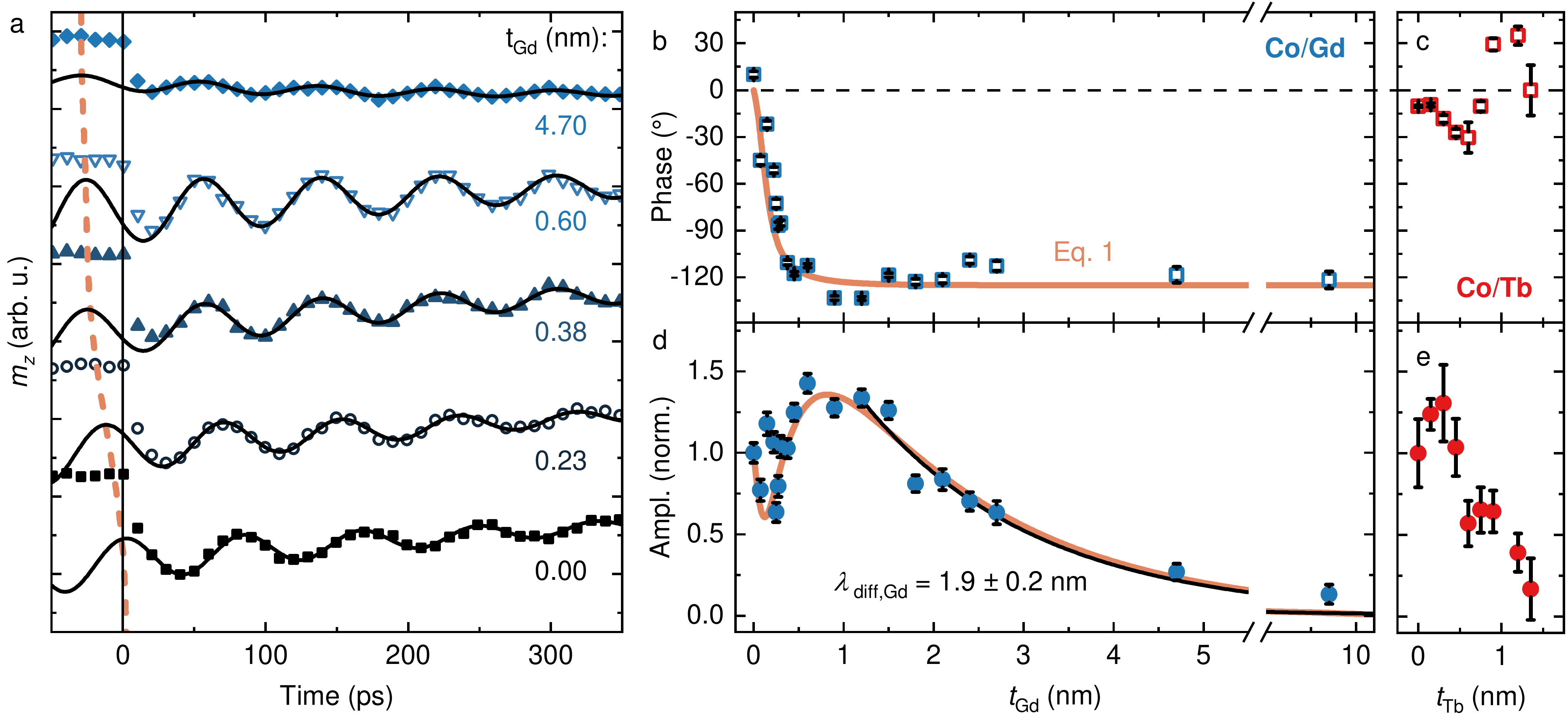}
\caption{\label{fig:2} 
(a) Measured FMR mode ($H_\text{ext}=100$ mT) for several Gd thicknesses, fitted with a damped cosine function. The orange dashed line indicates the FMR phase shift and the thickness calibration is discussed in Sec.\,1 of the Supplementary Information~\cite{Sup}.
(b,c) FMR phase as a function of Gd (b) and Tb (c) thickness.
(d,e) FMR amplitude as a function of Gd (d) and Tb (e) thickness, normalized to the amplitude without RE layer for both datasets.
The orange solid line in (b) and (d) represents the modeled behavior using eq. 1, and the black solid line in (d) is a fit with an exponential function.}
\end{figure*}

\section{Experimental results}
As a first experiment, we investigate the general behavior of the precessional dynamics for the three different generation layer configurations (Fig.\,\ref{fig:1}(a)).
In order to isolate the contribution of the RE layers, in which the magnetization decays exponentially away from the Co interface, their thicknesses need to be chosen carefully.
Previous work gives a typical lengthscale of the magnetization decay in Gd of 0.45 nm~\cite{lalieu2017aos}, with a similar value to be expected for Tb.
In Fig.\,\ref{fig:1}(c) and (d) we present typical measurements of the THz spin-wave and FMR mode respectively for the three different generation layer configurations, namely Co(1), Co(1)/Gd(0.53), and Co(1)/Tb(0.45).
Our results show that the composition of the generation layer significantly influences the spin-wave behavior, specifically the amplitude and phase.
To better understand this observation, we show representative modelled traces (See Sec\,V of the Supplementary Information~\cite{Sup}) of the magnetization response of Co and Gd upon fs laser excitation, in Fig.\,\ref{fig:1}(b).
The precessional behavior can then generally be understood with a simple picture where the generated spin current is assumed to be proportional to $\text{d}m/\text{d}t$~\cite{Choi2014, lichtenberg2022probing}.
Considering the antiferromagnetic coupling between Co and Gd (or Tb), the spin currents generated by the RE material and by Co should also have the opposite sign.
Therefore, we expect that the polarization of the total spin current changes sign (Fig.\,\ref{fig:1}(a)) when the RE material becomes the dominant contributor, which could explain the experimental observation of the large phase shift of the FMR mode. 
However, no such sign change is observed in the THz measurements, hinting at a less significant contribution of the RE material to the sub-ps spin current dynamics.
This is consistent with the expected magnetization dynamics plotted in Fig.\,\ref{fig:1}(b), which for Gd take place on the timescale of multiple ps, leading to a relatively slow spin current profile.
Additionally, we find a strongly reduced amplitude for precession excited with a Co/Tb generation layer.
This can be attributed to strong absorption of the spin current generated in Co, consistent with the large SOC in Tb~\cite{frietsch2020role}, as well as weak spin current generation in Tb.

\subsection{FMR mode}
To get a better understanding of the observed behavior, we systematically measure both the FMR and THz mode as a function of the Gd and Tb thickness in the generation layer.
First we discuss the results on the FMR mode, of which we show a selection of measurements for a Co(1)/Gd(X) generation layer in Fig.\,\ref{fig:2}(a).
To extract the phase and amplitude of the FMR mode, these measurements are fitted with a damped cosine function, indicated by the solid black lines, where the frequency is shared between all datasets.

We first discuss the FMR mode phase, of which we already observe a shift as a function of Gd thickness in Fig.\,\ref{fig:2}(a), as indicated by the dotted orange line connecting the shifted maxima of the fits. 
The phase we extract from the fits is plotted in Fig.\,\ref{fig:2}(b) and (c) for Co/Gd and Co/Tb respectively.
With the addition of only 1 nm of Gd, a phase shift of more than 130\textdegree{} is observed, whereas no consistent phase shift is measured when adding Tb.
This again indicates a strong contribution to the spin current from Gd, but a small to non-existent contribution from Tb.
For large thicknesses of Gd the phase remains constant, which can be explained by the paramagnetic state further away from the Co/Gd interface. 

Next, we focus on the amplitude of the FMR mode as a function of RE thickness, which is normalized to the amplitude for a pure Co generation layer and plotted in Fig.\,\ref{fig:2}(d) and (e) for Co/Gd and Co/Tb respectively.
This amplitude has been corrected for changes in light absorption using transfer matrix calculations~\cite{katsidis2002general}, as described in Sec.\,II of the Supplementary Information~\cite{Sup}.
For Gd, an initial dip and succesive rise of the amplitude is observed, followed by a gradual decrease. 
This behavior, in combination with the observed change of the FMR phase can be captured in a simple toy model where we assume the spin current contributions from Co and Gd excite two spin waves with different amplitudes and phases, but with the same frequency. 
As discussed previously, the Gd is magnetized due to exchange coupling with the Co, decaying at a characteristic length scale $\lambda_\text{mag}$ which determines its contribution to the total spin current.
Furthermore, the addition of Gd introduces a characteristic length scale $\lambda_\text{diff}$ over which spin information is lost due to spin flip scattering events, known as the spin diffusion length.
For now we assume that spin diffusion is independent of the magnetic state of Gd.
Writing the phase difference $\delta$ between the two excited precessions using Euler's formula, the FMR mode can then be described as
\begin{align}
    \label{Eq1}
    {A}_\text{FMR}~e^{i\phi} =\left(A_\text{Co} - A_\text{Gd}~e^{i\delta}\left(1-e^{-\frac{t_\text{Gd}}{\lambda_\text{mag}}}\right) \right)e^{-\frac{t_\text{Gd}}{\lambda_\text{diff}}}\, ,
\end{align}
where $A_\text{Co}$ and $A_\text{Gd}$ are the dimensionless amplitudes of the precessions excited by the magnetic volume of Co and Gd respectively.
These then result in a combined precession with amplitude ${A}_\text{FMR}$ and phase $\phi$.
Expressions for these two parameters can be derived, as is shown in Sec.\,III of the Supplementary Information~\cite{Sup}.
Now $A_\text{Co}=1$ can be fixed by normalization, and $A_\text{Gd}=3.2$ and $\delta=-140^\circ$ are chosen to match the maximum of the amplitude and saturation of the phase, respectively.
Within these constraints, valid values for the remaining parameters are found by manual adjustment, and are found to be $\lambda_\text{mag}=0.4\pm0.1$~nm and $\lambda_\text{diff}=2.0\pm0.2$~nm.
The amplitude of the Gd-excited precession $A_\text{Gd}$, when corrected for the expected equivalent magnetic thickness of Gd (0.45 nm), gives a Gd contribution that is approximately 7 times larger per nm than that of Co.
A complete understanding of this difference is outside of the scope of this work, but some factors of relevance are the differences in magnetization, the amount of magnetic moment lost during demagnetization, and the spin-wave excitation efficiency, which will be discussed later.
The value for $\lambda_\text{mag}$ closely matches the experimentally determined length scale for the loss of magnetization in Gd of $\sim0.45$nm~\cite{lalieu2017aos}.
We note that this length scale could also affect the rate of spin flip scattering, such that the parameter $\lambda_\text{diff}$ might not provide a full description.
Fitting the data for the FMR amplitude for thicknesses where Gd is expected to be paramagnetic (from 1.5 nm onwards) with an exponential function, indicated by the solid black line in Fig.\,\ref{fig:2}(d) results in a $\lambda_\text{diff,Gd}$ of $1.9\pm0.2$~nm, which has not been measured before to the best of our knowledge.
The close agreement between the two descriptions indicates that the magnetic state is not very relevant for spin flip scattering in these weakly magnetic systems.

The reasonable agreement between experiments and calculations indicate that the most important elements of this complex system are captured by this simple model.
However, the model can not capture the behavior we observe for a Co/Tb generation layer, as plotted in Fig.\,\ref{fig:2}(c) and (e).
Here we instead find only a rapid decrease of the amplitude with the addition of Tb.
This could indicate a very short diffusion length $\lambda_\text{diff}$ for Tb, which precludes any statements about the spin-current generation strength in Tb.
We attribute this discrepancy to a larger degree of scattering of the mobile electrons in Tb due to the high SOC compared to Gd~\cite{frietsch2020role}, leading to a loss of the spin information over shorter length scales.

\begin{figure}[t!]
\includegraphics[width=8.6 cm]{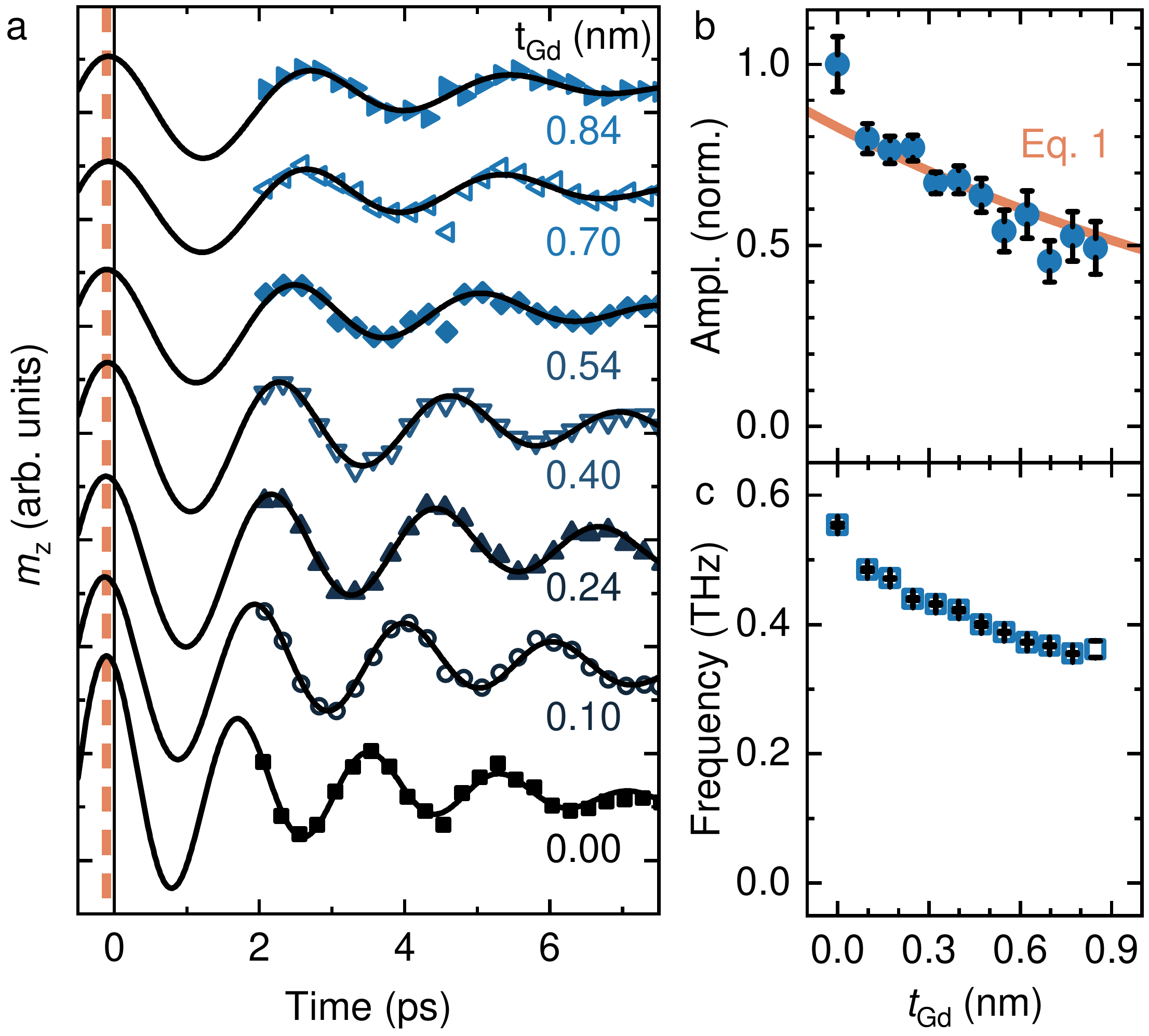}
\caption{\label{fig:3} 
(a) Measured THz mode for several Gd thicknesses, fitted with a damped cosine function. The orange dashed line indicates the spin-wave phase shift. 
(b) Spin-wave amplitude, orange line indicates a solution of Eq.\,1 using  $A_\text{Gd}=0$ and $\lambda_\text{diff,Gd}=1.9$~nm and 
(c) frequency as a function of Gd thickness.
} 
\end{figure}

\subsection{THz mode} 
We measure the THz spin-waves as a function of RE thickness to investigate the spin-current generation on the ps timescale, and show a selection of the measurements for a Co/Gd injection layer in Fig.\,\ref{fig:3}(a). 
Because THz mode excitation takes place on the same timescale as laser-induced demagnetization and spin-current generation, we disregard the first 2 ps of the data, in accordance with Ref.~\cite{lichtenberg2022probing}. 
We again used a damped cosine function to fit the data, indicated by the solid black lines, in order to extract the spin-wave amplitude, phase, and frequency. 
At around 1 nm, the signal-to-noise ratio is too low to extract spin-wave parameters reliably.
The amplitude of the spin waves, which is plotted in Fig.\,\ref{fig:3}(b), drops significantly with Gd thickness. 
Contrary to the behavior for the FMR mode, no initial rise of the amplitude due to a Gd contribution is observed.
We therefore show the expression for the amplitude derived from Eq.\,1 using $A_\text{Gd}=0$ and $\lambda_\text{diff,Gd}=1.9$~nm as the solid orange line in the figure, which is equivalent to an exponential decay describing only the spin diffusion due to Gd.
The good agreement indicates that Gd is not actively contributing to the excitation of the THz mode, which shows that this mode is excited by the Co-dominated fast component of the generated spin current \cite{Koopmans2010}. 
This is further confirmed by the observation that the spin-wave phase is independent of the Gd thickness, as indicated by the orange dashed line in Fig.\,\ref{fig:3}(a). 
The same measurements are repeated for Tb and presented in Sec.\,IV of the Supplementary Information~\cite{Sup}.
No significant difference between the THz spin-wave frequency and phase for the two materials is observed, again confirming the dominant role of Co in exciting these spin waves. 
However, a full analysis of the data is complicated by a more rapid decrease of the spin-wave amplitude with increasing Tb thickness, which could again be attributed to high spin-flip scattering due to SOC.

Contrary to our measurements of the FMR mode we observe a significant decrease of the THz spin-wave frequency for increasing Gd thickness, plotted in Fig.\,\ref{fig:3}(c), which is not predicted by our simulations (see Sec.\,VI of the Supplementary Information~\cite{Sup}).
We believe there are several mechanisms that could explain this observation, the foremost being a growth related change of the exchange stiffness \cite{eyrich2014effects, lalieu2017thz} and coupling between spin waves in the generation and absorption layers.
The mechanisms are discussed in detail in Sec.\,VII of the Supplemental Material~\cite{Sup} and predict the same direction of the shift of the THz spin-wave frequency.
However, a full understanding of this effect is beyond the scope of this work and requires further research.

\section{$s$-$d$ modelling}
To better understand the mechanisms governing our observations, as well as the underlying physics in general, we modeled the generated spin current in the synthetic ferrimagnetic generation layer using an extension of the $s$-$d$ model \cite{Gridnev2016, Cywinski2007, Manchon2012, Choi2014, Tveten2015, Shin2018, Beens2020}. 
This model describes the coupling of local spins, in this case the $3d$ and $4f$ electrons in the RE-TM ferrimagnet, to a system of itinerant spins ($s$ electrons). The latter system includes diffusive spin transport, similar to Refs.~\cite{Choi2014, Shin2018, Beens2020}. 
To model the experiments, we define a discretized material system consisting of a ferrimagnetic (Co/Gd) region and a nonmagnetic spacer layer (Cu). 
The Co/Gd bilayer is modeled in a layered manner, where the local Co and Gd concentration is sampled from a function that represents an intermixed transition from pure Co to Gd, similar to Ref.\,\cite{Beens2019intermixing}. 
This process is discussed in Sec.\,V of the Supplementary Material \cite{Sup}. 
Furthermore, the absorption layer is implemented as an ideal spin sink connected to the spacer layer, which is a valid assumption considering the experimental absorption layer thickness of 5 nm and the transverse spin diffusion length in Co of approximately 1 nm~\cite{lalieu2017thz}.
This allows us to calculate the absorbed spin current for varying composition of the Co/Gd layer. 
Using a linearized Landau-Lifshitz-Gilbert equation, including the anti-damping spin-transfer torque exerted by the absorbed spin current, we calculate the absolute phase shift of the excited homogeneous precession. 
Although the used model includes multiple assumptions that disqualify any precise quantitative statements~\cite{Sup}, it gives a complete description of the qualitative characteristics of the spin current and excited precession. 
Further details on the modeling, including all the used material parameters, are presented in Sec.\,V of the Supplemental Material~\cite{Sup}. 
\begin{figure}[t!]
\includegraphics[width=8.6 cm]{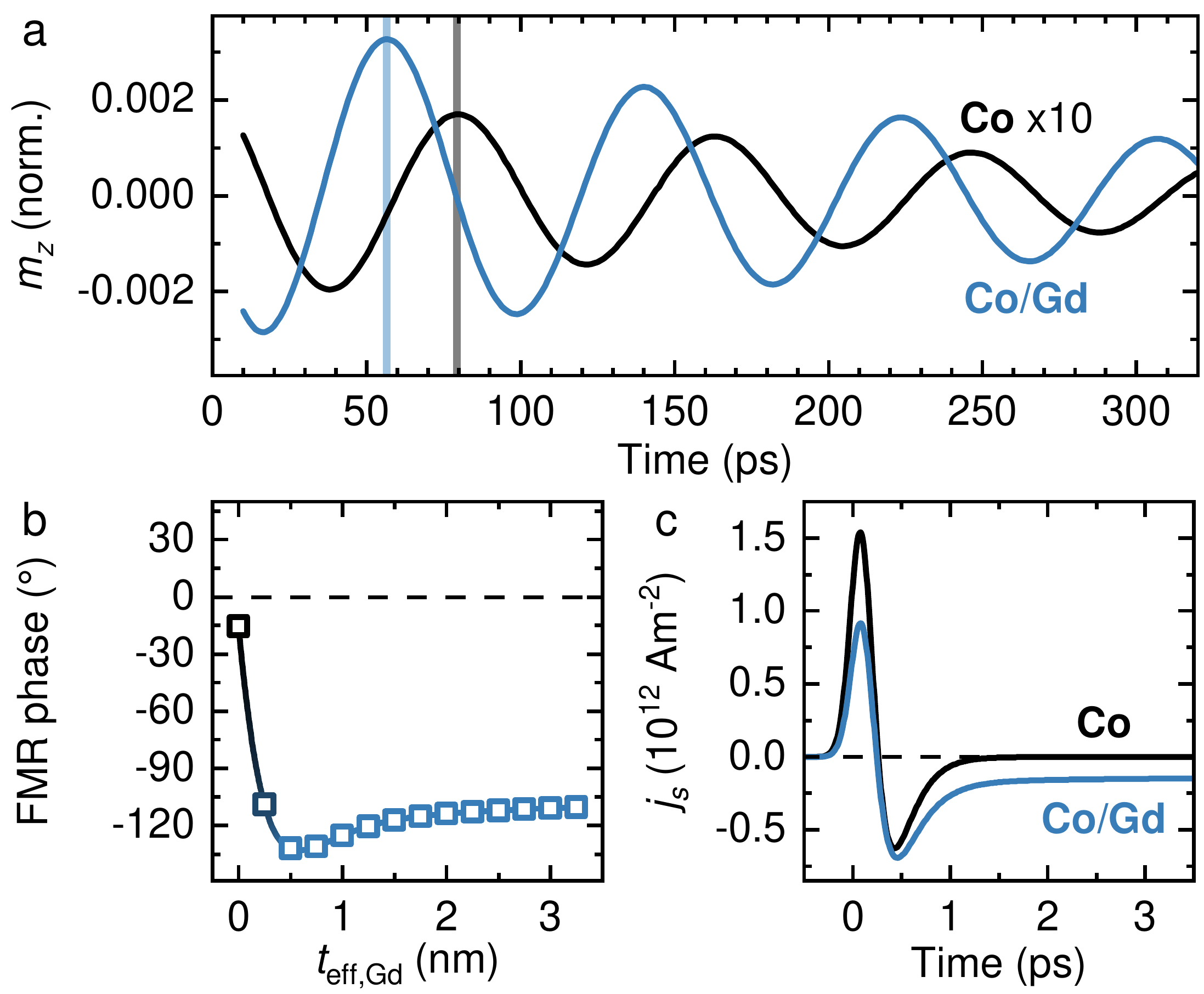}
\caption{\label{fig:4} 
(a) Calculated FMR mode excited by the spin current from a Co(1) (black) and Co(1)/Gd(2) (blue) generation layer. 
(b) Absolute phase of the excited FMR mode as a function of the effective thickness $t_\text{eff,Gd}$ of the (generation) Gd layer. 
(c) Absorbed spin current as function of time generated by a Co (black) and a Co/Gd (blue) generation layer. 
} 
\end{figure}

In Fig.\,\ref{fig:4}(a) we present traces of the modeled $z$-component of the FMR mode excited by a Co(1) and a Co(1)/Gd(2) generation layer, in black and blue, respectively.
Note that the amplitude of the FMR mode excited by pure Co has been multiplied by 10 for clarity, as it is significantly weaker than that excited by Co/Gd.
This is likely a result of the assumptions made in the model, such as the simplistic description of optical absorption and the parameters chosen for Gd, as the exact shape of the spin current pulse strongly affects the excitation efficiency.
Similar to the experimental results, our model shows a shift of the phase of the FMR mode when Gd is added to the generation layer.
In Fig.\,\ref{fig:4}(b) the absolute phase is plotted as a function of the effective Gd thickness $t_\text{eff,Gd}$, which for our purposes is analogous to the Gd layer thickness, and is further described in Sec.\,V of the Supplemental Material~\cite{Sup}.
The figure clearly shows a qualitative agreement with the experiments (Fig.\,\ref{fig:2}(d)), with a shift of approximately 130\textdegree{} upon adding a few monolayers of Gd.
The exact value of the phase shift can be explained by a combination of factors.
Naively, one would expect a 180\textdegree{} phase shift if only the Gd spin-current would excite dynamics once it starts to dominate the total spin current.
The actual shift however is lowered by the contribution of Co, as well as the relatively slow (Type II~\cite{Koopmans2010}) magnetization dynamics of the Gd, as will be demonstrated in the following.

To clarify the origin of the observed phase shift, we plot the absorbed spin currents for a pure Co(1) injection layer and a Co(1)/Gd(2) bilayer calculated using the $s$-$d$ model in Fig.\,\ref{fig:4}(c). 
The addition of Gd leads to changes in both the amplitude of the first, fast peak of the spin current, as well as a much longer negative tail.
Both of these can be understood by the demagnetization behavior of Gd, as plotted in Fig.\,\ref{fig:1}(b).
The reduction of the amplitude of the first peak of the spin current is caused by the initial, relatively rapid demagnetization of Gd.
At longer timescales, when the Co generated spin current is negligible, a long tail in the Co/Gd generated spin current is observed.
This is in turn caused by the secondary, slower demagnetization of Gd, and results in a spin current that is mostly polarized in the opposite direction.
This explains the abrupt phase jump in Fig.\,\ref{fig:2}(d) and \ref{fig:4}(b) for increasing Gd thickness.
Moreover, the long duration of this component of the spin current gives rise to a deviation of the phase shift from 180\textdegree{}.
This is to be expected, as the Gd demagnetization takes place over tens of ps, which is comparable to the precessional period ($\sim$ 80 ps).
The small increase of the phase from -130\textdegree{} to -110\textdegree{} that is observed in Fig.\,\ref{fig:4}(b) for a further increase of the Gd thickness is attributed to an additional slowing down of the Gd magnetization dynamics.
More specifically, we find that transfer of angular momentum from the Co, which speeds up the Gd magnetization dynamics near the interface between Co and Gd, has a smaller total effect when the Gd thickness is increased.

In Sec.\,VI of the supplementary information~\cite{Sup} we show calculated traces of the THz mode for increasing Gd thickness. 
Our calculations indicate a phase shift of up to 10\textdegree{} when adding a 2 nm thick Gd layer to a 1 nm thick Co layer. 
Contrary to previous work~\cite{lichtenberg2022probing} we were unable to resolve this potential phase shift, because the resolution is determined by the strength of the spin current and thus the total magnetic moment of the generation layer. In future research, Co/Gd multilayers could be used to study THz spin-wave generation in rare-earth transition-metal ferrimagnetic system in more detail. 

\section{Conclusion}
We have shown that examining the parameters of precessional modes excited by ultrafast optically generated spin currents can be a powerful tool to elucidate the behavior of these spin currents.
Synthetic ferrimagnets offer a novel platform to systematically investigate these phenomena.
The large discrepancy in spin current generation between Gd and Tb could shed new light on the relative difficulty of achieving AOS in systems containing only the latter as RE material.
Using spin-wave modes with THz frequencies has additionally allowed us to probe the high-frequency component of the generated spin currents.
Here we again find a link between the intrinsic speed of the magnetization dynamics and the behavior of the excited spin current.
This gives additional weight to the notion that angular momentum which is lost during demagnetization can be transferred to mobile spins~\cite{Choi2014}.
This notion also underlies the $s$-$d$ model, which we have successfully used to describe our experimental results.
Our experiments give new insight in the magnetization dynamics of rare-earth materials, which could prove critical in future spintronic memory devices.

\section*{Supplementary Material}
\beginsupplement
\setcounter{section}{0}
\section{Additional experimental details}

\noindent In this work, ultrafast laser-pulse induced dynamics are measured using pump-probe spectroscopy. 
A photo-elastic modulator is used to modulate the polarization of the probe, aligned such that the polarization oscillates between linearly and circularly polarized light at 50 kHz. 
To measure the pump-induced change of the magnetization after excitation, a mechanical chopper at around 69 Hz is used to modulate the pump. 
A dual lock-in scheme, locked to the two aforementioned frequencies, is used to measure the Kerr rotation induced in the sample. 
A more detailed description of the technique can be found in ref.~\cite{koopmans2000femtosecond}. 

In our samples, the Gd and Tb layers were deposited while moving a shutter over the sample, leading to a thickness gradient along the sample.
In order to determine the absolute thickness on this wedge an additional marker layer of Pt was deposited while keeping the shutter at the starting point of the path followed in the previous step.
The reflectivity of the sample as a function of position then exhibits a step where this marker layer starts, giving a calibration point for the gradient.
Due to shadowing effects the width of the step in this measurement is $\sim$0.3 mm, which corresponds to 0.3 nm of RE thickness.
Also due to imperfections in the positioning of the shutter, the exact starting point of the gradient can not be determined with greater accuracy than this 0.3 nm.
Within this range, we define 0 nm as the final measurement point along the sample before the behavior of the FMR or THz mode begins to change.
Note that the length of the gradient can be accurately determined, and the relative variation in thickness is accurate to within a few percent.

To accurately determine the exact moment of excitation, the so-called coherence peak is used~\cite{eichler1984coherence}, which is centered around the temporal pump-probe overlap. 
This is a non-magnetic effect, so it can be measured by measuring the demagnetization after saturation in opposite directions. 
Adding the two curves leaves only the non-magnetic effects that are independent of the magnetization direction, which in our case is only the coherence peak. 
Using this technique, the moment of excitation can be determined with an uncertainty in the order of 10 fs, necessary for phase measurements of the THz spin waves. 
A more detailed description of the measurements, as well as an example, can be found in ref.~\cite{lichtenberg2022probing}.

\newpage
\section{Transfer matrix calculations for optical absorption}
In Fig. 2b of the main paper we show the amplitude of the FMR mode for Gd thicknesses in the generation layer of up to 10 nm.
At these thicknesses the optical absorption starts to play a significant role, and needs to be taken into account when determining the actual amplitude.
As the film thickness increases, the absorbed laser energy also increases.
Therefore heating and thus demagnetization is greater for thicker Gd layers, potentially generating stronger spin currents.
To calculate the size of this effect, we perform transfer matrix calculations~\cite{katsidis2002general} using the \emph{tmm} Python package~\cite{byrnes2016tmmPython}.
Values for the refractive indices at the central laser wavelength (780 nm) are taken from literature~\cite{sopradatabase,akashev2019optical}.
\begin{figure}[h!]
\includegraphics[width=0.99\textwidth]{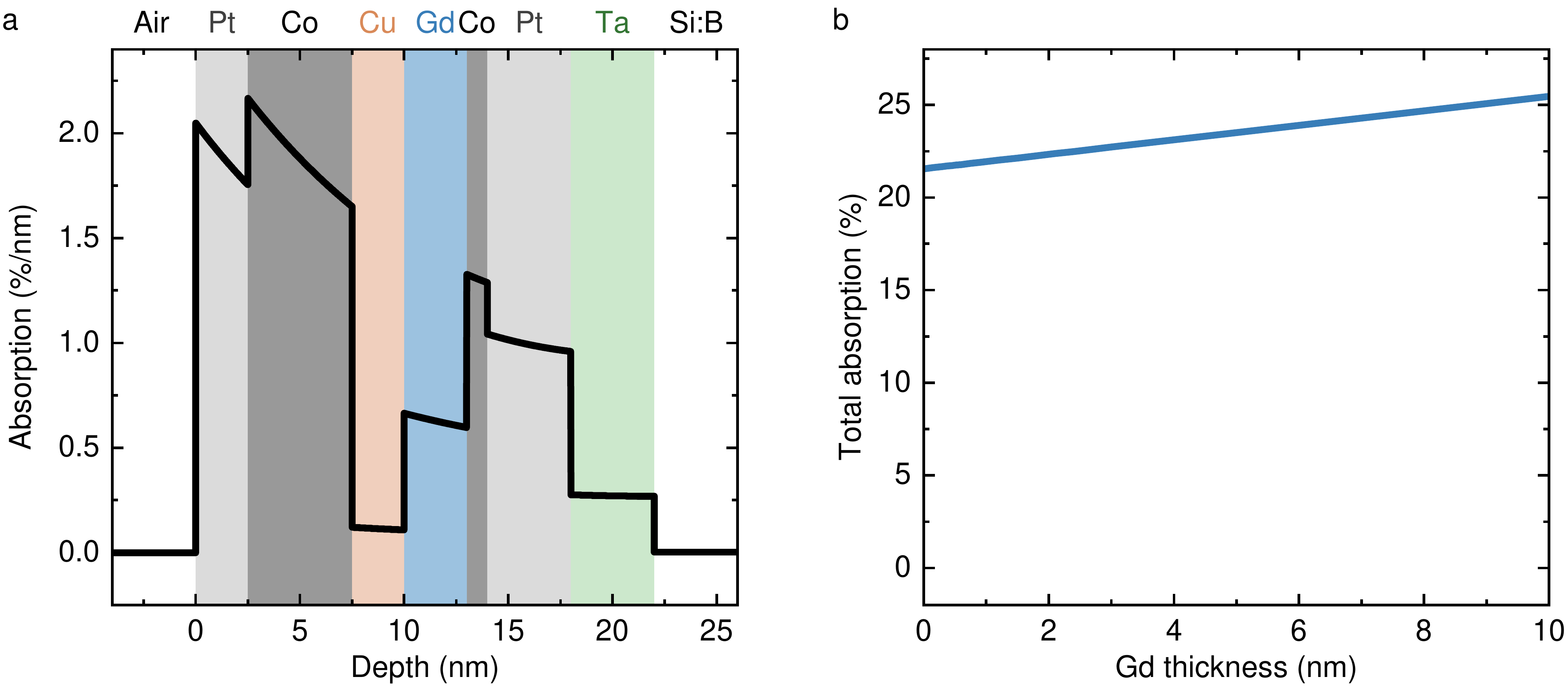}
\caption{\label{fig:II} 
(a) Instantaneous light absorption profile calculated with transfer matrix method for a Si:B(substrate)/{\allowbreak}Ta(4)/{\allowbreak}Pt(4){\allowbreak}/{\allowbreak}Co(1)/{\allowbreak}Gd(3)/{\allowbreak}Cu(2.5){\allowbreak}/{\allowbreak}Co(5){\allowbreak}/{\allowbreak}Pt(2.5) stack.
(b) Total absorption in the metallic layers as a function of the Gd thickness.
}
\end{figure}

We calculate the instantaneous absorption profile for the metallic stack used in the experiments (with $t_\text{Gd}$ = 3 nm), which we plot in Fig.\,\ref{fig:II}(a).
In metallic layers with thicknesses of only a few tens of nm, thermal equilibration between the layers proceeds on the timescale of mere hundreds of fs~\cite{xu2019single}.
As this is comparable to the demagnetization timescale in Co, and much shorter than this timescale in Gd~\cite{Koopmans2010}, the absorption in the entire metallic stack is the relevant parameter to determine the ultimate magnetization quenching.
Integrating the absorption profile over the thickness of the metallic layers then gives this total absorption.
Repeating this for the range of Gd thicknesses used in the main work gives the curve shown in Fig.\,\ref{fig:II}(b).
Here we indeed find an increase of the absorption with the Gd thickness.
To finally arrive at the data shown in Fig. 2b of the main paper, the measured amplitude of the FMR spin-waves is divided by the total absorption for each value of the Gd thickness.
\newpage

\section{Derivation of the phase and amplitude of toy model FMR mode} 
\noindent In the main text, we explain the Gd thickness dependence of the FMR amplitude and phase by decomposing the total FMR signal into a separate Co and Gd contribution, each with its own respective amplitude (${A}_\text{Co}$ and ${A}_\text{Gd}^\dagger={A}_\text{Gd}(1-\text{exp}(-t/\lambda_\text{mag}$)) and phase ($0$ and $\delta$). 
Using Euler's formula, this can be written as:
\begin{align}
\label{D1}
    {A}_\text{FMR} e^{i(\omega t+\phi)} = {A}_\text{Co} e^{i \omega t}+{A}_\text{Gd}^\dagger e^{i(\omega t+\delta)}\,. 
\end{align}

\noindent As the frequency $\omega$ is assumed to be the same for all contributions, we can multiply Eq,\,\ref{D1} with $e^{-i\omega t}$, yielding:
\begin{align}
\label{D2}
    {A}_\text{FMR} e^{i \phi} = {A}_\text{Co} + {A}_\text{Gd}^\dagger e^{i\delta}\,. 
\end{align}

\noindent First, we find an expression for ${A}_\text{FMR}$ by taking the complex absolute value of Eq.\,\ref{D2}.
\begin{align}
\label{D3}
    {A}_\text{FMR} &= \sqrt{|{A}_\text{Co} + {A}_\text{Gd}^\dagger e^{i\delta}|^2} \nonumber\\
    &= \sqrt{{{A}_\text{Co}}^2+{{A}_\text{Gd}}^2 \left(1-e^{-\frac{t}{\lambda_\text{mag}}}\right)^2+2 {A}_\text{Co} {A}_\text{Gd} \cos (\delta) \left(1-e^{-\frac{t}{\lambda_\text{mag}}}\right)}\,.
\end{align}
Next, an equation for the phase of the FMR mode is derived by taking the real and imaginary part of Eq.\,\ref{D2}:
\begin{equation}
\label{D4}
\left.
\begin{aligned}
  {A}_\text{FMR} &=\frac{{A}_\text{Co} + {A}_\text{Gd}^\dagger \cos \delta}{\cos \phi }\\
  {A}_\text{FMR} &= \frac{{A}_\text{Gd}^\dagger \sin \delta}{\sin \phi },
\end{aligned}
\right\}
\qquad \phi=\tan ^{-1}\left(\frac{\sin (\delta){A}_\text{Gd} \left(1-e^{-\frac{t}{\lambda_\text{mag}}}\right) }{{A}_\text{Co}+\cos (\delta ) {A}_\text{Gd} \left(1-e^{-\frac{t}{\lambda_\text{mag}}}\right)}\right)\, .
\end{equation}
Eq.\,\ref{D3} and \ref{D4} are used to describe the data presented in Fig.\,2b and d of the main text. 

\newpage
\section{THz spin-waves excited by Co/Tb}

\begin{figure}[h!]
\includegraphics[width=0.99\textwidth]{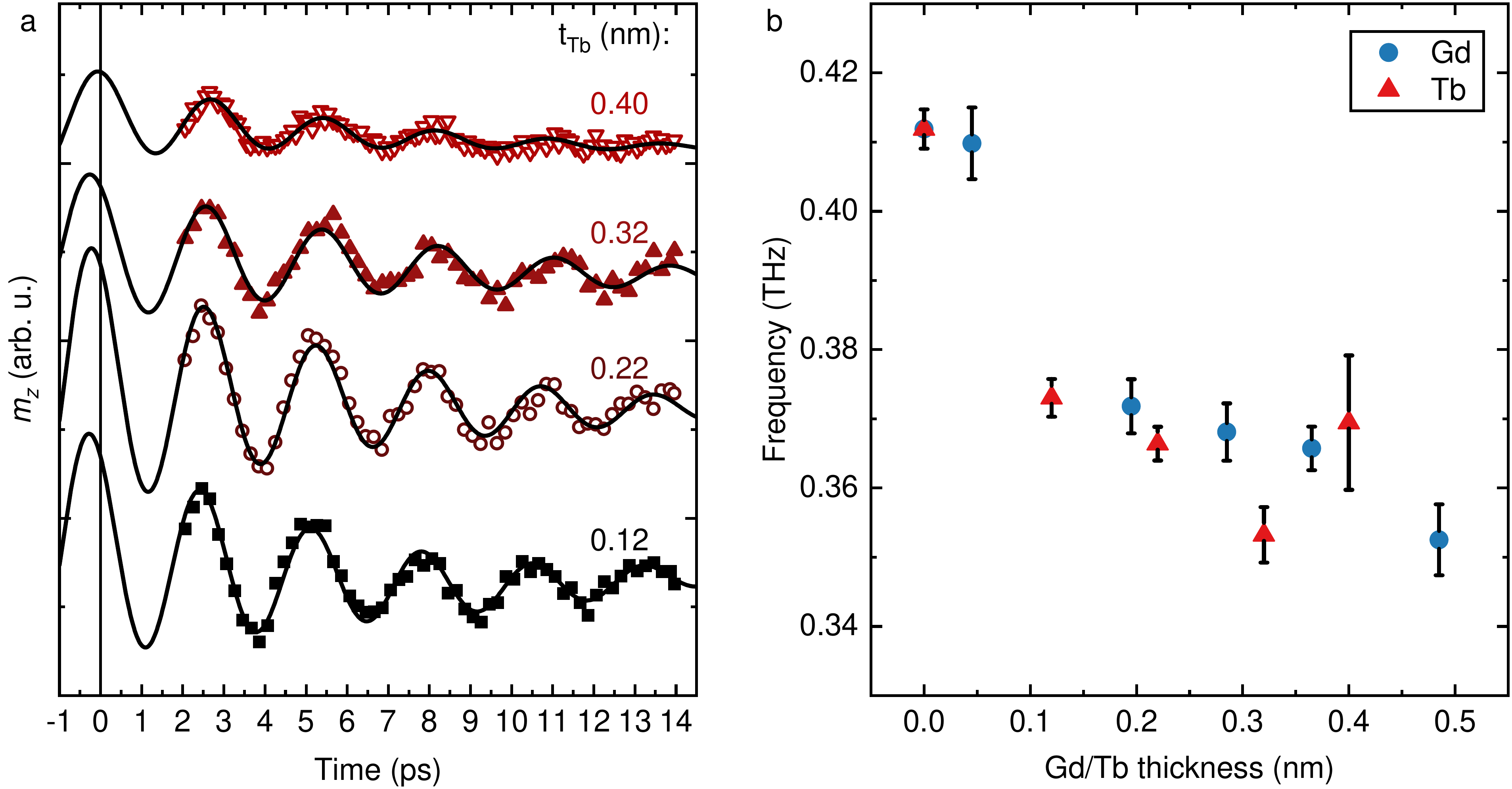}
\caption{\label{fig:V} (a) Spin wave measurements for the Co/Tb generation layer. (b) Extracted frequency as a function of Tb and Gd thickness. Data for Gd on a similar sample are inc;luded for comparison.}
\end{figure}

\noindent THz spin-wave measurements are performed for the Co/Tb generation layers as well, as presented in Fig.\,\ref{fig:V}(a). 
Due to strong attenuation of the measured signal for added Tb, only Tb thicknesses of up to 0.4 nm can be reliably considered. 
A decrease of the spin-wave frequency for increasing Tb thickness is observed (Fig.\,\ref{fig:V}(b)), similar to our observation for Gd, presented in the main text. 
We should note that these measurements were performed under slightly different growth conditions, leading to a slightly different absorption-layer thickness and thus a different spin-wave frequency. 
To be able to compare Gd and Tb, we performed the same experiments for a sample containing Gd, grown under the same conditions. 
As can be seen in Fig.\,\ref{fig:V}(b), the frequency does not depend on the RE material, indicating that THz spin-wave generation is dominated by the Co. 
Furthermore, the observed frequency differences between identical samples gives us a strong indication that the observed anomalous frequency behavior reported in the main text is related to growth. 

\newpage
\section{\emph{\lowercase{s}}-\emph{\lowercase{d}} modelling}

\subsection{The description of the magnetic stack} 

\noindent As mentioned in the main text, the generation layer is described as a system of layered CoGd alloys with a varying local Co concentration describing an intermixed Co/Gd bilayer \cite{beens2019comparing}. Furthermore, this Co/Gd bilayer is connected to a Cu spacer layer of $d=2.5 \mbox{nm}$. We define the interface between the generation layer and the spacer layer as $z=0$, where $z$ represents the out-of-plane spatial coordinate. Now the local concentration of Co at position $z$ is modeled by the function  
\begin{equation}
\label{SImaarten1}
x_\mathrm{Co} = \dfrac{1}{2} 
\bigg[
1-\mbox{erf}\Big(
(z-z_\mathrm{int})/w 
\Big)
\bigg],
\end{equation}

\noindent where $z_\mathrm{int}$ is the position of the the (intermixed)  Co/Gd interface. We set $w=0.5\mbox{ nm}$. This yields an intermixing region with a width in the order of nanometers, an order of magnitude larger than the discretization step size (lattice spacing $a$).   

Due to the intermixing region the layer thicknesses are not well defined. We define $\tilde{d}_\mathrm{Gd} = z_\mathrm{int}+a$, where $a$ is the lattice constant. $\tilde{d}_\mathrm{Gd}$ determines the total Gd in the system and plays the role analogue to a layer thickness. The total thickness of the system is then given $d=d_\mathrm{Co} + \tilde{d}_\mathrm{Gd}$, where $\tilde{d}_\mathrm{Co}=1\mbox{ nm}$ is set constant and plays a the role of a thickness for the Co. To be clear, using this approach $d_\mathrm{Co}$ and $\tilde{d}_\mathrm{Gd}$ do not precisely define a layer thickness, since we assume an intermixed interface. Nevertheless, for increasing value of $\tilde{d}_\mathrm{Gd}$ the relative amount of Gd increases in an monotonuous way.

\subsection{The calculation of the phase} 

\noindent To calculate the phase of the excited homogeneous precession in the absorption layer, we write down the Landau-Lifshitz-Gilbert-Slonzcewski equation for a homogeneous magnetization in the form \cite{tserkovnyak2005nonlocal}
\begin{equation}
\label{SImaarten2}
\dfrac{d\mathbf{m}}{dt} =
-\omega_0 \mathbf{m}\times \hat{z} +
\alpha \mathbf{m}\times \dfrac{d\mathbf{m}}{dt} 
+\mbox{\boldmath$\tau$}_{\mathrm{STT}},
\end{equation}
\noindent where $\omega_0$ is the angular FMR frequency and $\alpha $ the effective Gilbert damping.  $\mbox{\boldmath$\tau$}_{\mathrm{STT}}$ is the (anti-damping) spin-transfer torque exerted by the absorbed spin current. The latter is proportional to \cite{tserkovnyak2005nonlocal}
\begin{equation}
\label{SImaarten3}
\mbox{\boldmath$\tau$}_\mathbf{STT} 
\propto \mathbf{m}
\times (j_s(t)\hat{x} \times \mathbf{m}) . 
\end{equation}
\noindent We focus on small perturbations around the ground state $\mathbf{m}=\hat{z}$, assuming small transverse components in the $x$ (OOP direction) and $y$ direction $\delta m_x,\delta m_y \ll 1$. Furthermore, we define $\psi = \delta m_x + i\delta m_y $. Eqs.\,\ref{SImaarten2} and \ref{SImaarten3} then reduce  to  
\begin{equation}
\label{SImaarten4}
-i(1-i\alpha) \dfrac{d\psi}{dt} 
= \omega_0 \psi 
-i\tau_{\mathrm{STT},x}(t),
\end{equation}
\noindent which can easily be solved using the Green's function method. Assuming $|\alpha|^2\ll 1$, we have the Green's function
\begin{equation}
\label{SImaarten5}
G_0(t-t') \propto \theta(t-t') \exp(i\omega_0(1+i\alpha)(t-t') ) . 
\end{equation}
\noindent The transverse magnetization is expressed in terms of a convolution
\begin{equation}
\label{SImaarten6}
\delta m_x(t) \propto \mbox{Re}\Big\{G_0\ast j_s \Big\}. 
\end{equation}
\noindent Eq.\,\ref{SImaarten6} is used to determine the phase of the homogeneous precession excited by $j_s(t)$ as calculated by the method in the previous section. 

Extending the model to include the excited inhomogeneous modes can be accomplished by taking the following steps. The first step is to make the normalized magnetization $\mathbf{m}$ spatial dependent. The second step is to implement the exchange field in Eq.\ \ref{SImaarten2}, which is of the form $\mathbf{H}_\mathrm{ex} \propto A \nabla^2 \mathbf{m}$ with $A$ the spin-wave stiffness \cite{lalieu2017thz}. In analogy with Eq.\ \ref{SImaarten4}, the resulting equation of motion can be expressed in terms of the complex function $\psi(x,t)$, which is now dependent on spatial coordinate $x$. Importantly, it is assumed that all incoming (perpendicular) spins are absorbed at the interface. Now, the spin-transfer torque will appear as a boundary condition for the function $\psi(x,t)$. Finally, the function $\psi(x,t)$ is expanded in cosines $\cos(n\pi x/L) $ that represent the standing spin-wave solutions. In that way, an equation for every separate mode can be derived. It is straightforward to show that the contribution $\delta m_{x,n}$ of mode $n$ to the amplitude of the transverse magnetization is equivalently determined by 
\begin{equation}
\label{SImaarten7}
\delta m_{x,n}(t) \propto \mbox{Re}\Big\{G_n\ast j_s \Big\},
\end{equation}
with the response function $G_n(t)$ for mode $n$ 
\begin{equation}
\label{SImaarten8}
G_n(t-t') \propto \theta(t-t') \exp(i\omega_n(1+i\alpha_n)(t-t') ) ,
\end{equation}
where $\omega_n$ is the frequency of mode $n$ and given by $\omega_n=\omega_0+(A/\hbar) (n\pi/L)^2$ with $L$ the thickness of the absorption layer. Furthermore, $\alpha_n$ is the mode-dependent damping parameter. For simplicity, the latter is assumed to be independent of the mode and set equal to the Gilbert damping. 

\subsection{System parameters} 
\begin{table}[h!]
\centering
\caption{\label{tab:S1} The typical values used for the material parameters in the calculation of section .. .}
\begin{tabular}[t]{llr}
\hline \\ [-2.0ex] 
symbol & meaning &value  \\ 
\hline \\ [-1.5ex] 
$T_\mathrm{amb}$  & ambient temperature &$295\mbox{ K}$ \\ 
$\gamma $ & electronic heat capacity parameter & $ 2000 \mbox{ J\,m}^{-3}\mbox{\,K}^{-2} $  \\ 
$C_p$ & phonon heat capacity & $4\cdot 10^6\mbox{ J\,m}^{-3}\mbox{\,K}^{-1}$ \\ 
$P_0$ & absorbed pulse energy & $2\cdot 10^8\mbox{ J\,m}^{-3} $ \\ 
$ \sigma $ & pulse duration & $0.15 \mbox{ ps}$ \\ 
$\tau_D $ \footnotemark[1] & heat dissipation time scale & $100\mbox{ ps}$ \\ 
$a$ & lattice spacing  & $0.25 \mbox{ nm}$ \\ 
$T_{C,\mathrm{TM}} $ & TM Curie temperature & $1000\mbox{ K}$ \\ 
$T_{C,\mathrm{RE}} $ & RE Curie temperature & $292\mbox{ K}$ \\ 
$S_\mathrm{TM}$ & spin quantum number & $1/2$ \\
$S_\mathrm{RE}$ & spin quantum number & $7/2$ \\
$\mu_{\mathrm{at,TM}}$ & TM atomic magnetic moment & $2.0\,\,\mu_B$ \\
$\mu_{\mathrm{at,RE}}$ & RE atomic magnetic moment & $7.0\,\, \mu_B$ \\
$\rho_{sd}$ & $s$-$d$ coefficient Ref. \cite{Beens2020} & $1.0\mbox{ eV}$ \\ 
$\tau_{sd,\mathrm{TM}} $  & TM $s$-$d$ scattering time & $0.1\mbox{ ps}$ \\
$\tau_{sd,\mathrm{RE}} $  & RE $s$-$d$ scattering time & $20.0\mbox{ ps}$ \\
$ \tau_s $ & spin-flip scattering time (magnetic region) & $0.2\mbox{ ps}$ \\ 
$\tau_{s,\mathrm{N}} $ & spin-flip scattering time (nonmagnetic region)\cite{Shin2018} & $17\mbox{ ps}$ \\ 
$\sigma $ & conductivity (magnetic region) \cite{Shin2018} & $6.7\cdot 10^6\mbox{ S\,m}^{-1} $ \\ 
$\sigma_N $ & conductivity (spacer layer)\cite{Shin2018} & $39\cdot 10^6\mbox{ S\,m}^{-1}$ \\ 
$D$ & spin diffusion coefficient (magnetic region)\cite{Shin2018} & $250 \mbox{ nm}^2\mbox{\,ps}^{-1} $ \\ 
$D_N$ & spin diffusion coefficient (spacer layer)\cite{Shin2018} & $9500 \mbox{ nm}^2\mbox{\,ps}^{-1} $ \\ 
$g$ & interfacial conductance parameter & $0.4\cdot 10^6 \mbox{ m}^{-2}$ \\ 
$\omega_0/(2\pi) $ & FMR frequency &
$12 \mbox{ GHz}$\\ 
$\alpha$ & effective damping & 0.05 \\ 
$j_{\mathrm{Co-Gd}} $ & interatomic exchange constant & $2.0 \mbox{ meV} $ \\ 
\hline
\footnotetext{ The demagnetization traces imply that this time is extremely slow for the used stack, thereby set to an extremely long time. } 
\end{tabular}
\end{table}%

\noindent The values of the material parameters as used in the calculations presented in the main text are given in Table \ref{tab:S1}. We stress that although we use the terminology of Co/Gd components in the text, the material parameters presented here correspond to the values for a general TM-RE system. The latter is motivated by that the modeling only justifies qualitative statements, since the implementation required a large number of assumptions and approximations.

\clearpage
\section{Modelling THz spin waves using the \emph{\lowercase{s}}-\emph{\lowercase{d}} model}
\begin{figure}[h!]
\includegraphics[width=0.6\textwidth]{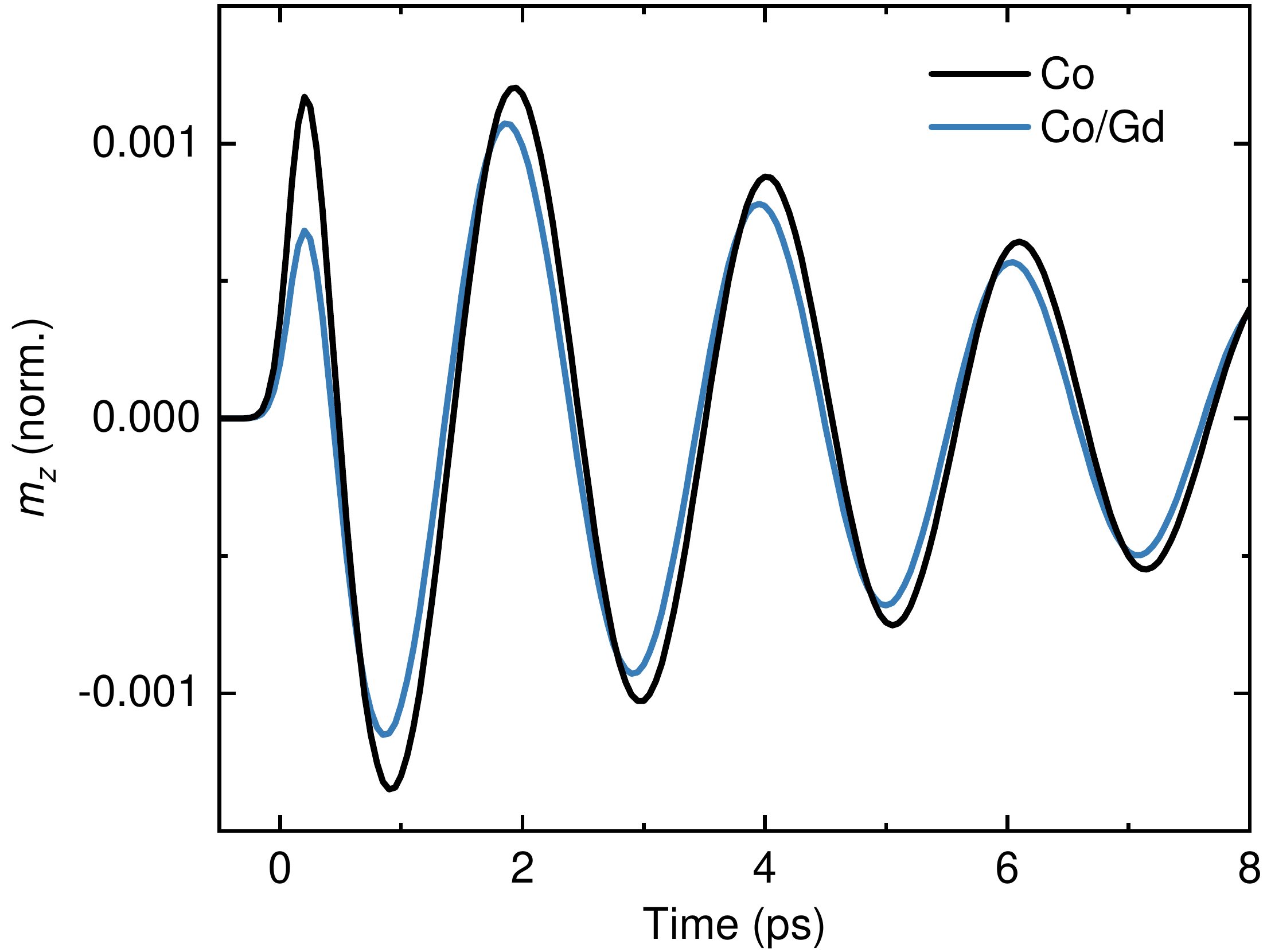}
\caption{\label{fig:IV} Calculated THz mode excited by the spin cur-
rent from a Co(1) (black) and Co(1)/Gd(2) (blue) generation
layer.}
\end{figure}

\noindent Using the $s$-$d$ framework discussed in Sec.\,IV of this, the THz response of the absorption layer is calculated. The results are plotted in Fig.\,\ref{fig:IV} for Co(1) (black line) and Co(1)/Gd(2) (Blue line).

Our calculations indicate a slight decrease of the amplitude, which in experiments we expect to be more pronounced due to spin diffusion in the Gd layer. 
Furthermore, a slight phase shift is expected.
The calculated shift is in the order of 60 fs, which corresponds to a phase shift of about 10\textdegree{}. This is too small to pick up in our measurements. 

\newpage
\section{Potential explanations for the THz spin-wave frequency shift}
\subsection{Growth Related frequency shift}

A growth-related change in the exchange stiffness scales linearly with the spin-wave frequency~\cite{vernon1984brillouin}. 
Eyrich \emph{et al.} demonstrated that the effective exchange stiffness of ultrathin ($<$ 10 nm) Co can drop significantly when alloyed with certain non-magnetic materials~\cite{eyrich2014effects}, and similar effects were later reported in Cu/Co bilayers due to intermixing at the interface~\cite{lalieu2017thz}. 
We postulate that due to the addition of a RE dusting below the Cu, subsequent growth becomes more disordered, leading to more intermixing and thus a lower exchange stiffness.
Although we expect the $\sim\!45\%$ reduction of the exchange stiffness required to explain the observed frequency shift completely is improbably large~\cite{eyrich2014effects, lalieu2017thz}, the order of magnitude of this effect is unknown for our system specifically and requires further research. 

\subsection{Interlayer coupling between two eigenmodes} 

We postulate that the frequency is significantly affected by the interlayer coupling of the simultaneous transverse magnetization dynamics in the generation layer and absorption layer. This coupling arises from the combination of spin pumping, interlayer spin transport, and the resulting spin-transfer torques. It will be maximized in case the eigenmodes of the absorption layer (the standing spin waves) and the eigenmodes of the generation layer (e.g., the exchange mode of the synthetic-ferrimagnetic multilayer) have a similar frequency. 

In order to reach a mathematical description of this concept, we formulate an  expression of the form of Eq.\ (\ref{SImaarten2}) for both layers separately. The noncollinear nature of the full magnetic stack is captured in the definitions of the effective fields within the distinct layers. In analogy with the previous subsection, the effective field within the absorption is pointing along the $z$ axis, and its magnitude is expressed in terms of the eigenfrequency $\omega_A$. For the generation layer, we define the effective field to be directed along the positive $x$ direction, with a magnitude determined by $\omega_G$. Additionally, we include the interlayer coupling terms. Specifically, the equation for the absorption layer includes the torque
\cite{tserkovnyak2005nonlocal}
\begin{equation}
\label{SImaarten9}
\mbox{\boldmath$\tau$}_\mathbf{\mathrm{interlayer-STT}} 
\propto 
\,\, g_A \, \mathbf{m}_A\times \bigg(\bigg( \mathbf{m}_G \times \dfrac{d\mathbf{m}_G}{dt} \bigg)\times \mathbf{m}_A \bigg), 
\end{equation}
where $\mathbf{m}_A$ describes the normalized magnetization of the absorption layer and $\mathbf{m}_G$ the normalized magnetization of the generation layer. Furthermore, the dimensionless factor $g_A$ is determined by the efficiency of spin pumping and the interlayer spin transfer, and mainly depends on the spin-mixing conductances. The expression describes an anti-damping spin-transfer torque mediated by the pumped spin current from the generation layer. A similar torque will be present in the equation for the magnetization dynamics in the generation layer, where the subscripts are interchanged $A \leftrightarrow G $. We linearize the set of equations using $\mathbf{m}_A= (\delta m_{A,x},\delta m_{A,y}, 1) $ and  $\mathbf{m}_G= (1,\delta m_{G,y}, \delta m_{G,z}) $. By collecting the linear terms, and switching to the frequency domain, we find the following equation for the eigenfrequency $\omega$
\begin{equation}
\label{SImaarten10}
((\omega_{A}+i\alpha_A \omega )^2 - \omega^2) 
((\omega_{G}+i\alpha_G \omega )^2 - \omega^2) 
-g_A g_G \omega^2 (\omega_{A}+i\alpha_A\, \omega ) (\omega_{G}+i\alpha_G \, \omega )=0,
\end{equation}
where $\alpha_A$ and $\alpha_G$ are the effective damping for the layer-specific eigenmodes. To purely extract the role of the effect discussed in this subsection, we remove the terms dependent on the damping $\alpha_A=\alpha_G\rightarrow 0$. By solving Eq.\ (\ref{SImaarten10}), it can be shown that the maximal shift of the frequency (compared to the layer-specific eigenfrequencies $\omega_A$ and $\omega_G$) is reached in case the eigenfrequencies are equal $\omega_A =\omega_G \equiv \omega_0 $. Then, the frequency is given by  
\begin{equation}
\label{SImaarten11}
\omega = \omega_0 \sqrt{1+\dfrac{g_A g_G}{2} \pm \sqrt{g_A g_G}  \sqrt{1+\dfrac{g_A g_G }{4}}  }. 
\end{equation}
\noindent For $\sqrt{g_A g_G}\ll 1$ this expression is equivalent to 
\begin{equation}
\label{SImaarten12}
\omega = \omega_0(1 \pm \sqrt{g_A g_G}/2), 
\end{equation}
indicating that the change is linear in $\sqrt{g_A g_G}$. 

Considering that the factors $g_A$ and $g_G$ are related to the effective damping (which for ultrathin films is strongly enhanced by spin pumping),  using an effective damping of $\alpha=0.1$ we estimate that the reduction of the frequency resulting from this effect is approximately $5\% $. Considering that the (not investigated) eigenmode of the ferrimagnetic generation layer may have a stronger damping, higher values for the frequency shift may be possible.  Although this mechanism can not explain the large frequency reduction as observed in the experiments, the order of magnitude of this results proofs that this contribution is nonnegligible.

\bibliography{Main_bib.bib}

\end{document}